# An all-dielectric photonic crystal with unconventional higher-order topology

Shiqiao Wu[1,2], Bin Jiang[1,2], Yang Liu[1,2], and Jian-Hua Jiang[1,2,3,†]

[1]*School of Physical Science and Technology, Soochow University, 1 Shizi Street, Suzhou, 215006, China*

[2]*Collaborative Innovation Center of Suzhou Nano Science and Technology, Suzhou, China*

[3]*Key Lab of Advanced Optical Manufacturing Technologies of Jiangsu Province and Key Lab of Modern Optical Technologies of Education Ministry of China, Soochow University, Suzhou 215006, China*

[†]Correspondence and requests for materials should be addressed to Jian-Hua Jiang via email: jianhuajiang@suda.edu.cn

## ABSTRACT

Photonic crystals have been demonstrated as a versatile platform for the study of topological phenomena. The recent discovery of higher-order topological insulators introduces new aspects of topological photonic crystals which are yet to be explored. Here, we propose a dielectric photonic crystal with unconventional higher-order band topology. Besides the conventional spectral features of gapped edge states and in-gap corner states, topological band theory predicts that the corner boundary of the higher-order topological insulator hosts a $\frac{2}{3}$ fractional charge. We demonstrate that in the photonic crystal such a fractional charge can be verified from the local density-of-states of photons, through the concept of local "spectral charge" as an analog of the local electric charge due to band filling anomaly in electronic systems. Furthermore, we show that by introducing a disclination in the proposed photonic crystal, localized states and a $\frac{2}{3}$ fractional spectral charge emerge around the disclination core, as the manifestation of the bulk-disclination correspondence. The predicted effects can be readily observed in the state-of-the-art experiments and may lead to potential applications in integrated and quantum photonics.

## I. INTRODUCTION

Topological insulators, which host gapped bulk states and robust gapless edge states [1-4], brought new concepts and ideas in photonics in the past decade. Topologically-protected photonic edge states can serve as robust waveguides, which have been demonstrated to be useful in integrated photonics [5-9], information transport [10-17], quantum photonics [18, 19], lasing [20-28] and exciton-polariton devices [29-31]. The ability to visualize the wavefunctions of the bulk and edge photonic states as well as the controllability of photonic systems make them a highly desirable platform for the study of topological phenomena.

Recently, it was predicted [30-39] and observed [40-68] that topological boundary states can emerge not only on the boundaries with $n-1$ dimensions, but also on the boundaries with $n-2$ and lower dimensions of $n$-dimensional ($n$D) topological insulators. Such exotic topological insulators are termed as higher-order topological insulators (HOTIs) [30-39]. For instance, a 2D HOTI hosts 1D edge states at the edge boundaries as well as 0D corner states at the corner boundaries [30-39]. The underlying mechanism is that due to the intricate role of the crystalline symmetry, the 1D edge states become gapped and hence can be regarded as emergent 1D insulators. At the corner boundaries between the edge boundaries, the 0D corner states emerge in the common band gap of the edge and bulk states due to edge or bulk band topology [30-68]. Such topologically-protected multidimensional boundary states beyond the bulk-edge correspondence introduce new degrees of freedom in the design of photonic states for wave-guiding, trapping and manipulation which may lead to potential applications in integrated photonics, quantum photonics and lasing [57-67]. However, such a discipline is still at its infant stage and yet to be explored. In particular, most existing studies focus on kagome and square lattice PhCs. Hexagonal PhCs which are very useful in PhC applications have rarely been studied for higher-order band topology.

In this work, we propose a hexagonal photonic crystal (PhC) which exhibits a topological band gap as the photonic analog of an unprecedented topological crystalline insulator. The unique band topology, as protected by the $C_6$ crystalline symmetry, is manifested first by the coexisting gapped edge states and in-gap corner states, indicating a photonic HOTI. From topological band theory, the photonic HOTI hosts a fractional corner charge of $Q_c = \frac{2}{3}$. We demonstrate using the first-principle simulations that such a fractional charge can be verified through the concept of local "spectral charge" as an analog of the local electric charge due to band filling anomaly in electronic systems. Physically, the "spectral charge" measures how many photonic modes exist in a local area in a given frequency range. Exploiting such a concept, we further show that disclinations, topological defects that disrupt the crystalline rotation symmetry, can induce a fractional spectral charge $Q_{dis} = \frac{2}{3}$ and trap localized photonic states around the disclination core. First-principle calculations give consistent results with the bulk-disclination correspondence picture predicted by topological band theory. The unique topological phenomena found in this work can be readily observed in the state-of-the-art photonic experiments and may offer potential applications in topological quantum photonics and topological lasing.

This paper is organized as follows: Sec. II focuses on the design of the PhC and its bulk topological indices. Sec. III studies the topological edge and corner states as well as the fractional corner charge. Sec. IV explores the manifestation of the bulk-disclination correspondence in the PhC. Sec.V demonstrates the disclination in the trivial photonic crystal. Sec.VI. shows the robustness of the disclination states. Sec. VII gives the conclusions and outlooks.

## II. PHOTONIC CRYSTAL AND BULK TOPOLOGICAL INDICES

We propose a hexagonal PhC with six dielectric cylindrical rods in each unit-cell [see Fig. 1(a)] to realize the above topological effects. We focus on the lowest few photonic bands of transvers-magnetic harmonic modes. Despite that our PhC looks a bit similar to the PhC proposed by Wu and Hu (denoted as "Wu-Hu's PhC" hereafter) [11], their topological indices and properties are distinct. First, there are only two bands below the topological gap in our PhC, whereas there are three bands below the topological gap in Wu-Hu's PhC. Moreover, the band symmetry representations and the topological indices are distinct for the two PhCs as elaborated below.

The photonic band gaps discussed here are analogs of topological crystalline insulators protected by the six-fold rotation ($C_6$) symmetry. In this context, the topological indices of the band gap can be deduced from the symmetry indicators of the bands below the gap. Using the theory in Ref. [39], the topological indices are given by

$$\chi^{(6)} = (\chi_M, \chi_K). \qquad (1)$$

Here, the symmetry indicators are $\chi_M = \#M_1^{(2)} - \#\Gamma_1^{(2)}$ and $\chi_K = \#K_1^{(3)} - \#\Gamma_1^{(3)}$, respectively. The symbol $\#\Pi_l^{(n)}$ denotes the number of bands with the $C_n$ rotation eigenvalue $e^{i2\pi(l-1)/n}$ ($l = 1, \ldots, n$) below the band gap at a high-symmetry point

$\Pi = \Gamma, M, K$. The symbol $\chi_M$ ($\chi_K$) thus stands for the change of the symmetry representations between the $M$ ($K$) and the $\Gamma$ points. For instance, $\chi_M$ represents the parity inversion between the $\Gamma$ and $M$ points, as generalized parity-inversion [1] (or generalized Fu-Kane [2]) indices. According to the symmetry eigenvalues at the high symmetry points [as illustrated in Figs. 1(b) and 1(c)], we find that our PhC has $\chi^{(6)} = (0, -2)$, whereas the Wu-Hu's PhC has $\chi^{(6)} = (-2, 0)$. Moreover, Wu-Hu's PhC has a nontrivial second Stiefel-Whitney number $\nu = 1$ [i.e., $4m + 2$ ($m$ is an integer) odd-parity Bloch states at all time-reversal invariant momenta (the $\Gamma$ point and the three $M$ points)], whereas our PhC here has a trivial second Stiefel-Whitney number $\nu = 0$ (i.e., $4m$ odd-parity Bloch states at all time-reversal invariant momenta), according to Ref. [69].

Geometrically, the key difference between our PhC and Wu-Hu's PhC is that in our PhC, the dielectric rods are aligned along the line from the corners to the center of the unit-cell [see the brown dashed line in the inset of Fig. 1(a)]. In comparison, in Wu-Hu's PhC the dielectric rods are aligned from the edge-centers to the center of the unit-cell [see the green dotted line in the inset of Fig. 1(a)].

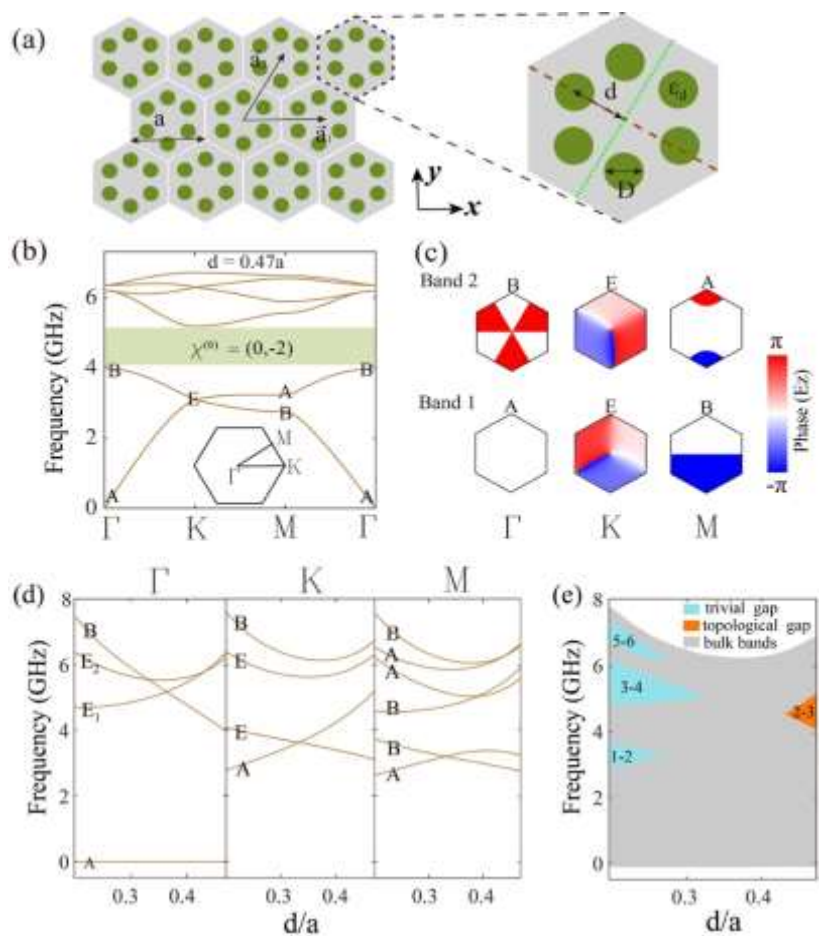

FIG. 1. (a) Schematic illustration of the 2D hexagonal PhC. Inset shows the zoom-in structure of the unit-cell. Gray region is air, while the green dots denote the dielectric rods with a diameter $D = 4$mm and a relative permittivity $\varepsilon_d = 24$. The lattice vectors of the hexagonal lattice, $\vec{a}_1$ and $\vec{a}_2$, are depicted. The lattice constant is $a = 25$mm, while the distance between the center of a rod and the unit-cell center, $d$, is tunable. (b) Photonic band structure for $d = 0.47a$. The photonic band gap (green region) has a nontrivial topological index of $\chi^{(6)} = (0, -2)$. The little group representations (i.e., A, B and E) are labeled on the band structure. The phase profiles for the electric field $E_z$ of the corresponding Bloch states are shown in (c). (d) Evolutions of the first six bands at the high symmetry points, $\Gamma$, $M$ and $K$, with the geometry parameter $d$. (e) Evolutions of the bulk bands (frequency ranges) and the band gaps with the parameter $d$.

The evolution of the lowest six photonic bands with the geometry parameter $d$, i.e., the distance between the rod center and the unit-cell center, is presented in Figs. 1(d) and 1(e). From Fig. 1(e), one can see that only the band gap between the second and the third bands is topological, whereas the other band gaps are all trivial (i.e., these band gaps have trivial indices $\chi_M = \chi_K = 0$). We find that the topological band gap is finite only when $d > 0.418a$.

Although the results in this paper do not depend on the lattice constant (since the Maxwell's equations are scale-invariant), most of the calculations here focus on the situation with $a = 25$mm and $d = 0.47a$ (unless specified as other values), while the relative dielectric constant and the diameter of the rods are $\varepsilon_d = 24$ and $D = 4$mm, respectively. For these parameters, we find a bulk band gap ranging from 4 GHz to 5.15 GHz. The band-gap-to-mid-gap ratio is as large as 25% which is advantageous for applications exploiting the edge and corner states.

### III. TOPOLOGICAL EDGE AND CORNER STATES AND THE FRACTIONAL CORNER CHARGE

As a consequence of the nontrivial topological band gap, topological edge and corner states emerge in finite structures. To demonstrate this, we first consider a structure which is finite along the $y$ direction, but periodic along the $x$ direction [see Figs. 2(a) and 2(b)]. Hard-wall boundary conditions are imposed at the two terminating edges in the $y$ direction which are simulated using the perfect electrical conductor (PEC) boundaries (i.e., metallic boundaries for far-infrared and microwave photons) [Fig. 2(b)]. Calculation indicates that the edge states are gapped and only one branch of the edge states emerge in the bulk band gap [Fig. 2(a)]. The dispersion of the edge states tends to be flat instead of gapless and dispersive. Such edge states are quite different from the known photonic edge states in the literature [5-17,68]. The amplitude and phase profiles of the electric field $E_z$ as well as the distributions of the Poynting vectors, showing in Fig. 2(b), agree with the time-reversal symmetry for the two edge states with opposite wavevectors.

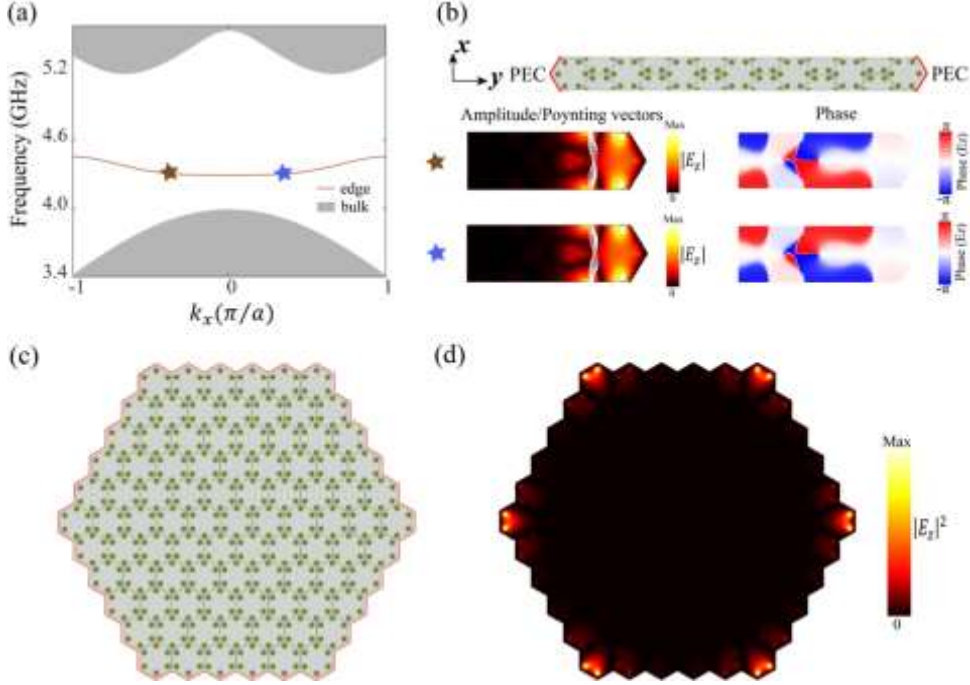

FIG. 2. (a) Photonic dispersions at the bulk, edge and corners. Both the bulk and edge dispersions are calculated using the structure illustrated in the upper panel of (b). The corner frequency is calculated from the finite structure illustrated in (c). Upper panel of (b): the structure for the study of the edge states which is finite along the $y$ direction but periodic in the $x$ direction. PEC boundaries are imposed at the left and right ends of the structure. Lower panels of (b): the amplitude and phase profiles of the electric field $E_z$ for the two edge states labeled by the stars in (a). (c) Illustration of the finite structure used to study the corner states. (d) Distribution of the summed electric field intensity for all the corner states. There are six degenerate corner states [see Fig. 3(a)]. In all calculations, a thin layer of air with the width $0.08a$ is set between the PEC boundary and the PhC.

When the PhC is finite in all directions, as depicted in Fig. 2(c), both the edge and corner states emerge. There are six nearly degenerate corner states emerging in the common band gap of the bulk and the edge states [see Fig. 3(a)]. Each of them is localized at one of the corner boundary, as indicated by the distribution of the summed electric field intensity $|E_z|^2$ for all the corner states [Fig. 2(d)].

In addition to the edge and corner states, topological band theory predicts that in finite systems, a fractional electric charge appears at the corner boundary due to the filling anomaly of the occupied bulk bands in electronic systems [39]. The fractional part of the corner charge, $eQ_c$, is connected to the bulk topological indices through the following relation [39],

$$Q_c = \left(\frac{1}{4}\chi_M + \frac{1}{6}\chi_K\right) \mod 1. \tag{2}$$

The above equation states that the fractional corner charge is determined by the bulk band topology. In electronic systems, the fractional corner charge is directly related to the local electric charge. However, photons are neutral particles which lack such a property. Nevertheless, it has been shown that through the concept of "spectral charge", the fractional corner charge can still be measured in bosonic systems [70].

To verify the fractional corner charge in photonic systems, we calculate the spectral charge for each unit-cell in the structure illustrated in Fig. 3(c). Specifically, the spectral charge for the $p$-th unit-cell in the structure is defined as follows,

$$Q_p = \int_0^{f_{gap}} df \int_{\substack{p-th \\ U.C.}} d\vec{r}\, \rho(f, \vec{r}), \tag{3}$$

where $f_{gap}$ is a frequency in the common band gap of the bulk and edge states, $\rho(f, \vec{r})$ is the LDOS of photons ($f$ is the frequency and $\vec{r}$ is the position vector). The integration over the coordinates is performed for the region of the $p$-th unit cell. The above equation is an analog of the electric charge for the $p$-th unit-cell by band filling up to the Fermi energy in the bulk band gap in electronic systems. For topologically trivial insulators, for each unit-cell $Q_p$ is equal to the number of bands below the band gap [39], which is consistent with the picture that each unit-cell contributes $Q_p$ modes to form the bulk bands below the trivial band gap. However, for topological crystalline insulators, fractional spectral charges can appear due to band filling anomaly [39].

We calculate the photonic LDOS and obtain the spectral charges for all the unit-cells. The photonic spectral charge for each unit-cell can be obtained by integrating the photonic local density-of-states (LDOS) $\rho(f, \vec{r})$ up to a frequency in the band gap, $f_{gap}$, as indicated by Eq. (3). The LDOS is calculated through the photonic eigenstates. If we denote the wavefunction of the $i$-th photonic eigenstate as $\psi_i$ and the frequency of the eigenstate as $f_i$, then the LDOS can be written as,

$$\rho(f,\vec{r}) = \sum_i \frac{\Gamma}{\pi[\Gamma^2+(f-f_i)^2]} |\psi_i(\vec{r})|^2 \qquad (4)$$

Here, $\Gamma$ is a parameter used to model the Lorentz broadening of the eigenstates. In our calculation, $\Gamma$ is set to be sufficiently small to converge the calculation. The normalized photonic wavefunction is given by

$$|\psi_i(\vec{r})|^2 = \varepsilon(\vec{r})|E_{z,i}(\vec{r})|^2 \qquad (5)$$

where $\varepsilon(\vec{r})$ is the relative permittivity and $E_{z,i}(\vec{r})$ is the rescaled electric field of the $i$-th photonic eigenstate which satisfies $\int d\vec{r}\, \varepsilon(\vec{r})|E_{z,i}(\vec{r})|^2 = 1$. The spectral charge is then given by Eq. (3), where the integration over the coordinates is performed for the region of the $p$-th unit cell.

Based on the above calculation method, we find that in the bulk region (gray), the spectral charge is nearly 2 for each unit-cell in Fig.3(b). In comparison, in the edge region (blue), the spectral charge is close to 1 for each unit-cell, while in the corner region (red), the spectral charge is close to $\frac{2}{3}$.

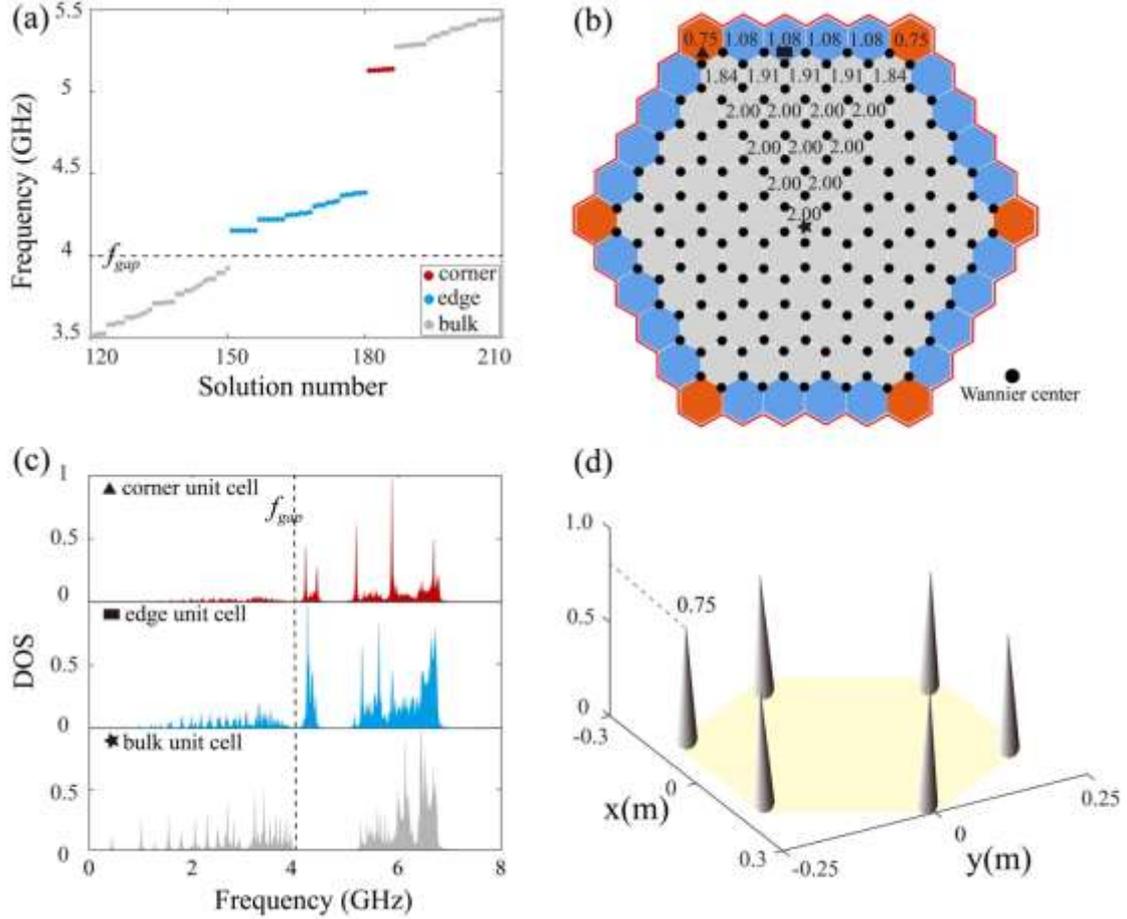

FIG. 3. (a) The photonic spectrum for the finite PhC structure illustrated in Fig. 2(c). Only the eigenstates in and around the bulk band gap are shown. (b) The spectral charges for various unit-cells. The gray region denotes the bulk. The blue region denotes the edges, while the red region denotes the corners. (c) Local density-of-states (DOS) for photons in three different types of unit-cells labeled by the triangle (corner unit-cell), rectangle (edge unit-cell), and the star (bulk unit-cell). Integrating the photonic LDOS up to the frequency $f_{gap} = 4$GHz gives rise to fractional spectral charges at the corner boundaries. (d) Illustration of the distribution of the fractional corner charges in a finite sample. In all calculations, a thin layer of air with the width $0.08a$ is set between the PEC boundary and the PhC.

The above spectral charges can be understood through the Wannier centers. In our photonic system, there are only bulk states below the frequency $f_{gap} = 4$GHz. These bulk modes are pictorially represented by the Wannier centers away from the edge and corner boundaries. In each bulk unit-cell, there are six Wannier centers locating at the corners of the unit-cell. Each Wannier center is shared by three neighboring unit-cells, thus contributing 1/3 spectral charge to one of these unit-cells.

Therefore, each bulk unit-cell has a spectral charge 2, i.e., there are two photonic modes in each bulk unit-cell. These two modes interact with the modes in other unit-cells and form the two bulk Bloch bands below the band gap.

In comparison, there are three Wannier centers in an edge unit-cell. The other three Wannier centers in the edge unit-cell are obstructed by the edge boundary and become edge states. For a corner unit-cell, there are only two Wannier centers associated with the bulk, while the other four Wannier centers are obstructed by the boundary and become edge and corner states. As a consequence, an edge unit-cell has a spectral charge 1, while a corner unit-cell has a spectral charge 2/3.

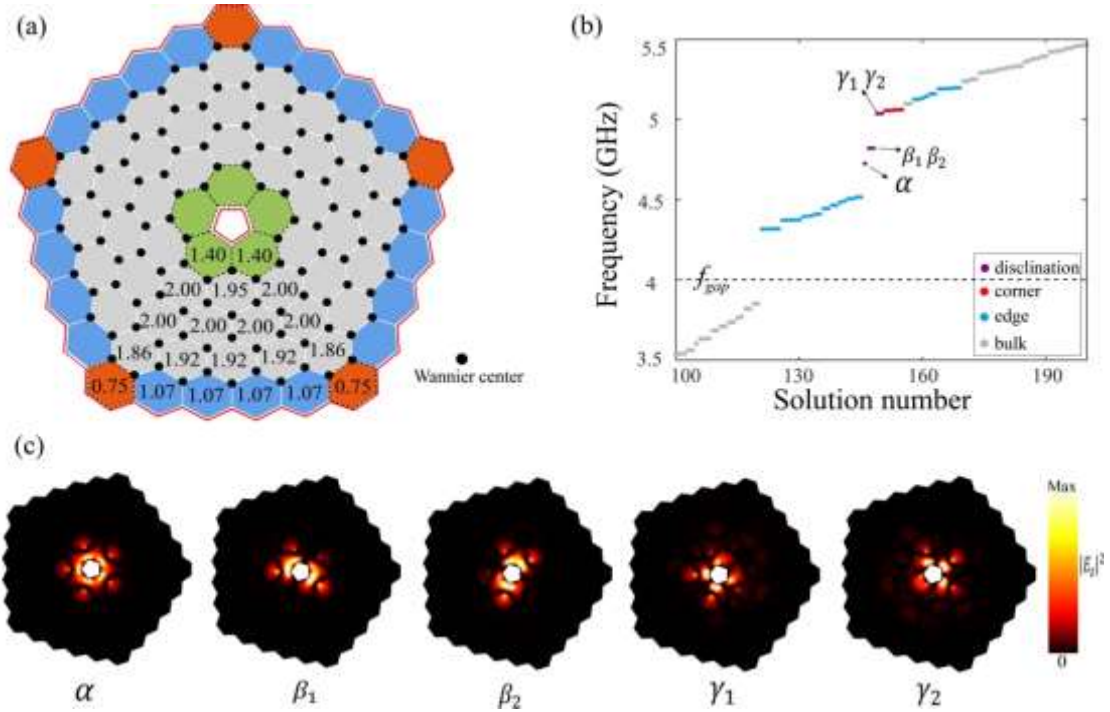

FIG. 4. (a) Spectral charges for various unit-cells in a finite disclination structure from the first-principle calculations. Integrating the calculated photonic LDOS up to the frequency $f_{gap} = 4$GHz gives the spectral charges presented in the figure. (b) Spectrum of the photonic eigenstates for the finite-sized disclination structure. (c) The five localized states bound to the disclination core. In all calculations, a thin layer of air with the width $0.08a$ is set between the outside PEC boundary and the PhC and another layer of air with the width $0.03a$ is set between the inner PEC boundary and the PhC.

Our first-principle calculation gives a spectral charge 0.96 for an edge unit-cell which is close to the theoretical value 1. On the other hand, each corner unit-cell has a spectral charge 0.75 which is not far away the theoretical value 2/3. Note that the charge has to be fractionalized to integer times of 1/3 or 1/2, not to other fractional values, according to Eqs. (2) and (6). Therefore, the small deviations of the numerical value of the corner (and disclination) charge from the theoretical value do not cause a problem in identifying the correct fractional corner or disclination charges. The slight deviation of the spectral charges calculated using the first-principle methods from the theoretical spectral charges essentially originates from the fact that photons in PhCs do not follow strictly the tight-binding theory. Nevertheless, the fractional corner charge is still revealed approximately from the first-principle calculations.

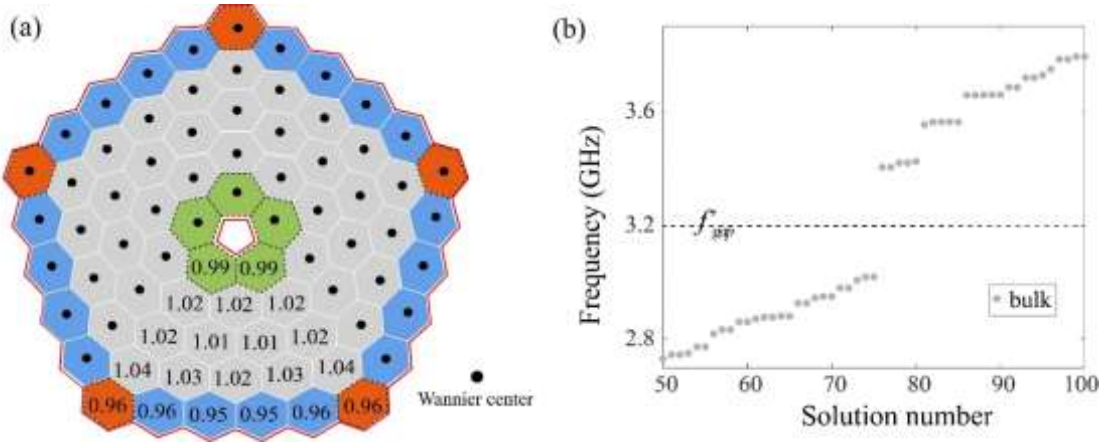

FIG.5. (a) Wannier center distributions in the disclination structure for the trivial PhC with $d/a = 0.25$. No fractional spectral charge is accumulated due to the localized position of Wannier center is at the center of each unit cell. (b) Photonic spectrum of the disclination structure for the trivial PhC. There is no disclination state. Only bulk states are found. In all calculations, a thin layer of air with the width $0.08a$ is set between the outside PEC boundary and the PhC and another layer of air with the width $0.03a$ is set between the inner PEC boundary and the PhC.

## IV. BULK-DISCLINATION CORRESPONDENCE

We now study the bulk-disclination correspondence in our higher-order topological PhC. The bulk-disclination correspondence predicts that in topological crystalline insulators, the fractional charge bound to a disclination with a Frank angle $\Omega$ is determined by the symmetry indicators of the Bloch bands below the band gap [71-73],

$$Q_{dis} = \frac{\Omega}{2\pi}\left(\frac{3}{2}\chi_M - \chi_K\right) \bmod 1, \tag{6}$$

where $\Omega$ is the Frank angle. The disclination structure in Fig. 4(a) has a Frank angle $\Omega = -\frac{2\pi}{6}$. This disclination structure has a fractional charge of $Q_{dis} = \frac{2}{3} \bmod 1$ bound to the disclination core. The bulk-disclination correspondence is manifested in both such a fractional disclination charge and localized states bound to the disclination core. The fractional disclination charge can also be understood via counting the number of the Wannier centers. Each unit-cell in the disclination region [green in Fig. 4(a)] has four bulk Wannier centers, which gives 4/3 spectral charge per unit-cell. The other two Wannier centers are obstructed by the disclination boundary and thus form localized disclination states. The five unit-cells close to the disclination core give in total a fractional spectral charge of 2/3.

To visualize the bulk-disclination correspondence in our PhC, we calculate the eigenstates and the LDOS of photons for the disclination structure in Fig. 4(a). The photonic spectrum [Fig. 4(b)] shows that there are five disclination states in the bulk band gap, in addition to the edge states. By integrating the photonic LDOS up to a frequency in the band gap, $f_{gap} = 4\text{GHz}$, we calculate the spectral charges for all the unit-cells. Due to the five-fold rotation symmetry of the disclination structure, we present the spectral charges only for part of the disclination [see Fig. 4(a)]. Again, each bulk unit-cell has a spectral charge close to 2, while each edge unit-cell has a spectral charge approximately 1. The disclination unit-cells have a spectral charge of 1.4 which is close to the theoretical value of 4/3.

### V. Disclination in the trivial photonic crystal

We now study the disclination structure for the trivial PhC. Specifically, we study the PhC with $d/a = 0.25$ and consider the photonic band gap between the first and the second bands. For the band gap, the topological indices are trivial (i.e., $\chi_M = \chi_K = 0$), and the Wannier center of the first band lies at the center of the unit cell [see Fig. 5(a)].

We calculate the spectral charge for each unit-cell by integrating the photonic LDOS from zero up to a frequency $f'_{gap} = 3.2$GHz in the concerned photonic band gap [see Fig. 5(b)]. The results in Fig. 5(a) indicates that all the spectral charges are close to 1. There is no signature of charge fractionalization in the disclination structure. Note that the charge has to be fractionalized to integer times of 1/3 or 1/2, not to other fractional values, according to the bulk-disclination correspondence (i.e., Eq. (6)). Moreover, there is no disclination state. All the eigenstates are bulk states.

## VI. Robustness of the disclination states

In this section, we study the robustness of the disclination states against two kinds of disorder: (i) defects that preserve the five-fold rotation symmetry of the disclination (i.e., at the center of the disclination core); (ii) defects that break such rotation symmetry (i.e., away from the center of the disclination core). The defect is realized by inserting a dielectric rod with a radius of 2 mm but with varying relative permittivity. To facilitate the discussions, we remove the PEC boundary in the disclination core. Such a set-up creates a hollow core in the disclination. As a consequence, a defect photonic mode emerges at the disclination core in the trivial PhC. In the meanwhile, there are still five disclination states localized around the disclination core.

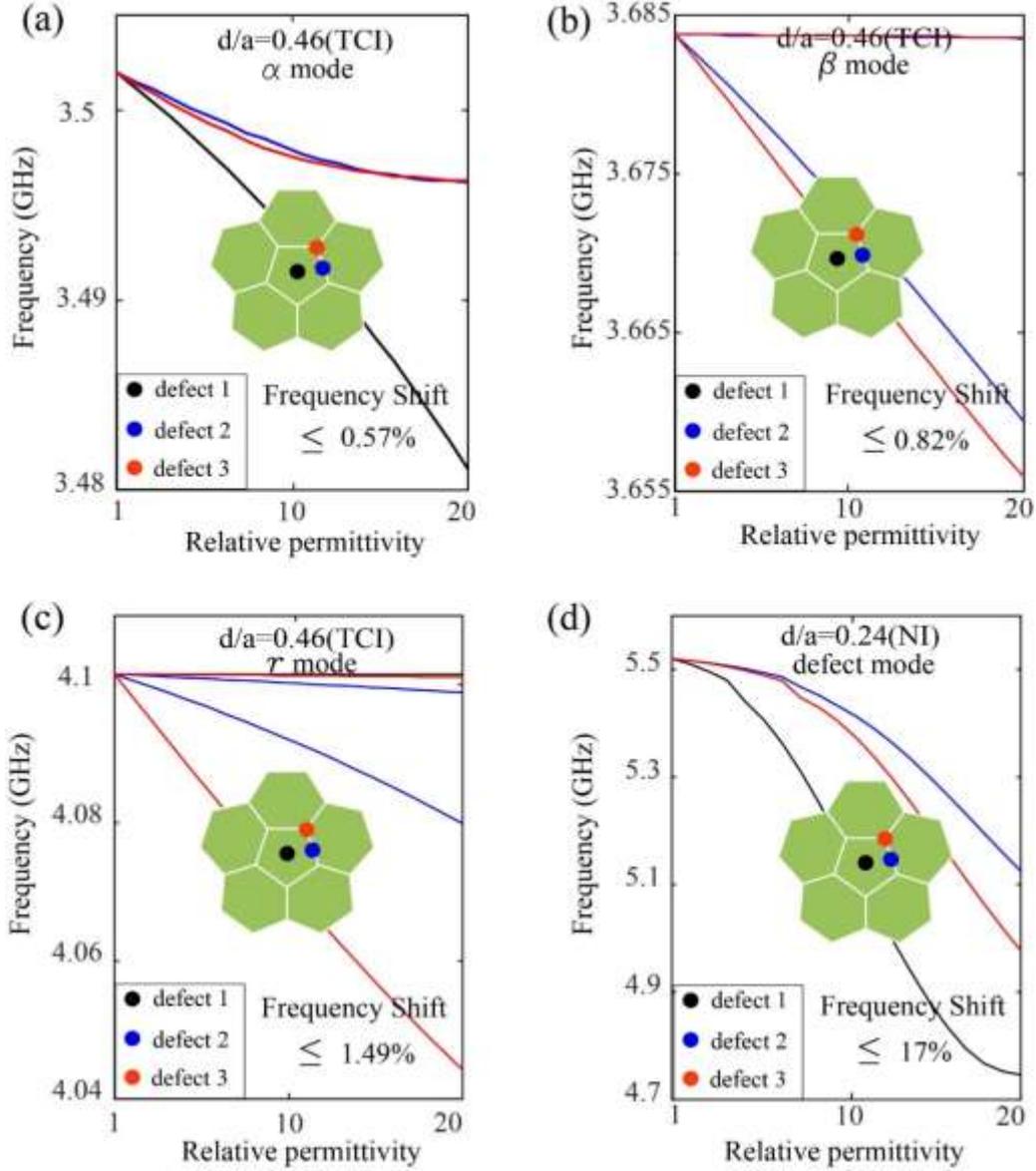

FIG.6. Frequency shift of the disclination and defect states when disclination structure contains an additional dielectric rod near the disclination core. The location of the defect rod is indicated by the black, blue and red dots in the insets. The radius of the defect pillar is 2 mm. We study the frequency-shifts of the disclination and defect states as functions of the relative permittivity of defect rod. Frequency of disclination states ($\alpha, \beta$ or $\gamma$) and defect state as functions of relative permittivity of dielectric rod for three different defects is summarized in Figs.6 a-c and d, respectively.

We then compare the robustness of the frequencies of the disclination states and the defect state against the same disorder. Specifically, we study the shift of the frequencies of the disclination states and the defect mode as functions of the relative permittivity of the inserted dielectric rod (acting as a defect) which ranges from 1 to 20 in the simulation. In order to show the robustness of the disclination states, we comparatively study two cases: the disclination structure formed by the photonic TCI with $d/a = 0.46$, and the disclination structure formed by the photonic NI with $d/a = 0.24$. The former has five disclination states, while the latter has a defect mode localized at the hollow disclination core. These two cases are adopted because they have the photonic band gaps with nearly the same band-gap-to-mid-gap ratio of 20%. The position of the defect rod is indicated in the insets in Fig. 6 by the black, blue and red dots, respectively. We compare the robustness of the frequencies of the disclination states and the defective state in responses to the same disorder. And the frequency of the same disclination state ($\alpha, \beta$ or $\gamma$) or defect state changes as functions of relative permittivity of dielectric rod and is summarized in the same figure. The results are presented in Fig.6 where three different configurations of the disorder are studied for both the TCI and NI. Specifically, Figs. 6(a) - 6(c) show, respectively, the effects of the defect pillar on the frequencies of the disclination states in the TCI. And Fig.6(d) shows the effects of the defect pillar on the frequencies of the defect state in the NI. It is seen that the frequency-shifts of the disclination states in the TCI are much smaller than the frequency-shift of the defect state in the NI. Thus, the frequencies of the disclination states in the TCI are more robust than the frequency of the defect state in the NI. These cases corresponding to defect configurations that break the five-fold rotation symmetry of the original disclination structure. Thus, the double-degeneracy in the $\beta$ and $\gamma$ states are lifted. Nevertheless, the resilience of the frequencies of the disclination states in the TCI over the frequency of the defective state in the NI remains visible.

## VII. CONCLUSIONS AND OUTLOOK

We propose a hexagonal photonic crystal with unconventional higher-order topology. In addition to the conventional spectral features of HOTIs, i.e., gapped edge states and in-gap corner states, the unique band topology here gives rise to a fractional charge of 2/3 at the corner boundaries which is confirmed by the first-principle calculations through the concept of spectral charges. The spectral charges measure the number of photonic modes within a local area (e.g., a unit-cell) for all the bulk states below the band gap. We also show that the bulk-disclination correspondence leads to a fractional spectral charge of 1/3 at the disclination core. Besides, we find that there are five localized states bound to the disclination which is robust against disorders when compared with the conventional defect mode in PhCs. In contrast, the above phenomena disappear in trivial photonic band gaps.

The localized states bound to disclinations can be used as photonic cavity modes which are robust against disorders. Such robust subwavelength cavities are useful in integrated photonic systems as well as for quantum photonics. In addition, these cavity modes can also be exploited for lasing as demonstrated in recent works. The fractional charges at the corners and disclination cores can be used to control the LDOS of the bulk photonic states. Our work may inspire future studies on similar phenomena and their applications in photonic and optoelectronic systems.

## ACKNOWLEDGEMENTS

This work is supported by the National Natural Science Foundation of China under Grant Nos. 12074281 and 12047541. J.-H. Jiang is supported by the Jiangsu specially-appointed professor funding, and a project funded by the Priority Academic Program Development of Jiangsu Higher Education Institutions (PAPD).

**Disclosures.** The authors declare no conflicts of interest